\documentclass[letterpaper, twoside]{JHEP3}
\usepackage{epsfig}

\usepackage{amsmath}

\title{Entropy creation inside black holes points to observer complementarity}

\author{Gavin Polhemus \\
	JILA, Box 440, University of Colorado, Boulder CO 80309, U.S.A. \\
	Poudre High School, 201 Impala Drive, Fort Collins, CO 80521, U.S.A. \\
	E-mail: \email{gavin.polhemus@colorado.edu}
}
\author{Andrew J S Hamilton \\
	JILA, Box 440, University of Colorado, Boulder CO 80309, U.S.A.\\
	Dept.\ Astrophysical \& Planetary Sciences, Box 391, \\
	University of Colorado, Boulder CO 80309, U.S.A.\\ 
	E-mail: \email{Andrew.Hamilton@colorado.edu}
}
\author{Colin S Wallace \\
	Dept.\ Astrophysical \& Planetary Sciences, Box 391, \\
	University of Colorado, Boulder CO 80309, U.S.A. \\
	E-mail: \email{Colin.Wallace@colorado.edu}
}

\newcommand{\abs}[1]{\left| #1 \right|}
\newcommand{\infinity}{\infty}

\newcommand{\sub}[1]{_{\textrm{#1}}}

\abstract{
Heating processes inside large black holes can produce tremendous amounts of entropy.  Locality requires that this entropy adds on space-like surfaces, but the resulting entropy ($10^{10}$ times the Bekenstein-Hawking entropy in an example presented in the companion paper) exceeds the maximum entropy that can be accommodated by the black hole's degrees of freedom. Observer complementarity, which proposes a proliferation of non-local identifications inside the black hole, allows the entropy to be accommodated as long as individual observers inside the black hole see less than the Bekenstein-Hawking entropy.  In the specific model considered with huge entropy production, we show that individual observers do see less than the Bekenstein-Hawking entropy, offering strong support for observer complementarity.
}

\keywords{Black Holes, Black Holes in String Theory}

\begin{document}

\section{Introduction}\label{introduction}

In a companion paper, we found that the entropy
produced inside a charged black hole can exceed the entropy released in the black hole's evaporation by many orders of magnitude \cite{Wallace:2008zz}.
If locality holds for black holes, then entropy is additive on space-like slices, and the excess entropy example in \cite{Wallace:2008zz} violates the second law of thermodynamics.\!%
\footnote{Entropy in this paper is always the entropy used in practical thermodynamic calculations, found by ignoring all quantum entanglements beyond some coarse graining scale.}
However, locality has been called into question in the case of black holes.  Susskind et al.\ \cite{Susskind:1993if} have argued that unitarity requires non-local identifications between degrees of freedom inside the black hole and degrees of freedom outside, an idea known as black hole complementarity.
The interior and exterior offer different, complementary views of the same degrees of freedom. Therefore, entropy on the interior does not add to entropy outside, and
the excess entropy example does not violate the second law.

By identifying the interior and exterior degrees of freedom, black hole complementarity limits the number of degrees of freedom inside the black hole to the number seen from the outside perspective.  The entropy computed in the excess entropy example is far too large (by a factor of $10^{10}$) to be accommodated by those degrees of freedom.
A stronger form of complementarity, observer complementarity, proposes additional non-local identifications across every observer horizon \cite{Banks:2001yp, Dyson:2002pf, Parikh:2002py}.  Observer complementarity predicts that while the entropy on space-like surfaces inside the black hole may exceed the Bekenstein-Hawking entropy, the entropy seen by individual observers cannot.  We show that even with the huge entropy seen in the excess entropy example, individual observers see less than the Bekenstein-Hawking entropy.  This is a strong confirmation of observer complementarity.

\section{Black hole thermodynamics supports black hole complementarity}\label{thermo}

Even when there is no entropy created inside the black hole, certain choices of space-like time slices can show apparent violations of the second law, as explained below.%
\footnote{Locality is the quantum-field-theory proposition that space-like separated to operators commute.  Commuting operators identify distinct degrees of freedom.  The entropy associated with distinct degrees of freedom is additive, so locality requires entropy to add on space-like slices.}
Black hole complementarity resolves this conundrum by proposing nonlocal identifications between the interior and exterior of the black hole, making the interior entropy redundant.  When the interior entropy is ignored in accordance with black hole complementarity, the second law is restored.
Unitarity arguments already support the idea of black hole complementarity \cite{Susskind:1993if}, but the thermodynamic argument below will clarify some of the ideas involved in the discussion of observer complementarity which begins in section \ref{entropy}.

To an outside observer, black holes behave like any other black body.  The entropy of the black hole is the Bekenstein-Hawking entropy, $S\sub{BH}$, which is equal to one quarter of the horizon area in Planck units.  When entropy falls into a black hole the horizon area always increases enough to prevent a violation of the second law.   In general, the entropy that goes into the formation of the black hole is much less than $S\sub{BH}$, so black hole formation is a thermodynamically irreversible event \cite{Bousso:2002ju}.

A black hole's temperature is given by $dE = T\,dS$, where the energy $E$ of the black hole is its mass.
Like any warm black body, a black hole radiates.
Black holes are unusual in that their temperature increases as their energy decreases, so they will radiate all of their energy in a finite time, ending with an explosion.
Black hole evaporation is thermodynamically irreversible---the entropy released during evaporation is somewhat higher than the black hole's entropy, $S\sub{evap} \approx \frac{3}{2} S\sub{BH}$.%
\footnote{The ratio of $S\sub{evap}$ to $S\sub{BH}$ depends on the number and polarizations of particle species radiated during evaporation.  Inclusion of three massless neutrinos and their antiparticles gives $S\sub{evap} \approx 1.6187\cdot S\sub{BH}$ \cite{Page:1983ug}.  We now know know that at least some of the neutrinos have mass which, while small, exceeds the Hawking temperature of a typical astronomical black hole.  Assuming all neutrinos are massive, $S\sub{evap} \approx 1.4848\cdot S\sub{BH}$ \cite{PagePrivate2009}.}
As viewed by outside observers, entropy increases at every stage from formation to evaporation, in accordance with the second law.

The above story of increasing entropy is vague about where the black hole's entropy resides and how it is reemitted as Hawking radiation.
To add some clarity, one could follow the entropy as it flows through the black hole space-time, adding the entropy on space-like slices, as dictated by locality.
The entropy should increase from one slice to the next in accordance with the second law.  
However, this integration can lead to apparent violations of the second law, even when no additional entropy is produced inside the black hole.  

To see the apparent violation of the second law, consider the space-like ``nice slice'' shown in figures \ref{NiceSlice} and \ref{Penrose} \cite{Giddings:2007ie}. This slice avoids Planck scale densities and curvatures, and is intersected by all of the entropy that falls into the black hole and by much of the Hawking radiation.  An explicit case of a violation occurs if, for example, the entropy of the material forming the black hole is approximately $\frac{5}{6}$ of the Bekenstein-Hawking entropy, and the slice intersects the horizon when a third of the mass has radiated away.
\begin{equation}
	S\sub{int} + S\sub{ext} 
	\approx \frac{5}{6}S\sub{BH} + \frac{3}{2}\left[1-\left(\frac{2}{3}\right)^{2}\right]S\sub{BH}
	= \frac{5}{3}S\sub{BH}\,.
\end{equation}
This exceeds $S\sub{evap}\approx \frac{3}{2}S\sub{BH}$, leading to a decrease in entropy when the black hole evaporates, in violation of the second law.

\DOUBLEFIGURE[t]{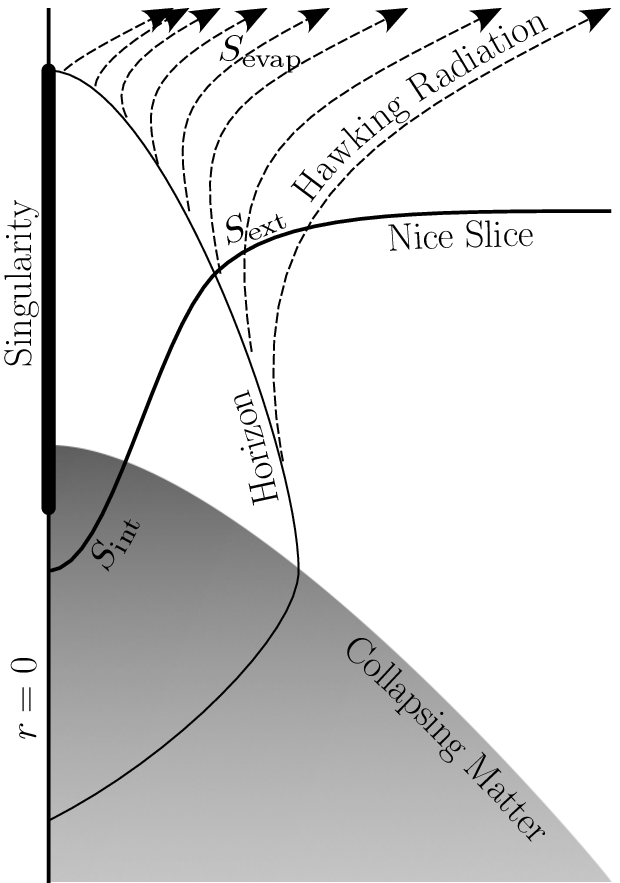}{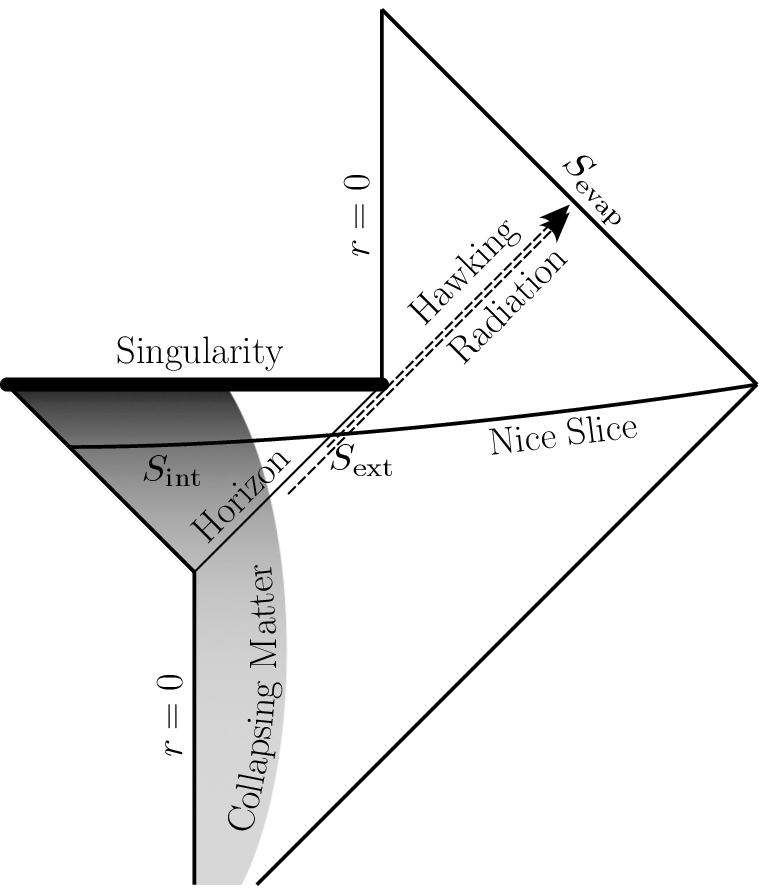}
{The total entropy on the space-like nice slice can exceed the final entropy, $S\sub{evap}$.  The rate of evaporation is greatly exaggerated.\label{NiceSlice}}
{The nice slice, which looks rather convoluted in figure 1,
is perfectly natural in the Penrose diagram.\label{Penrose}}

The source of this violation is, of course, the entropy inside the black hole.  The problem can be avoided entirely by taking the total entropy to be only the external entropy.%
\footnote{To get everything exactly right, there are subtle issues involved in counting the external entropy. The external entropy includes the black hole's thermal atmosphere.  The external region must end a tiny bit outside the actual horizon and entropy may need to be attributed to the boundary to account for the gap.  All of this accounting involves degrees of freedom outside the horizon.\cite{Thorne:1986iy}} 
Black hole complementarity, which was originally proposed to protect unitary black hole evolution, provides a rationale for counting only the outside entropy.  Black hole complementarity asserts that the interior of the black hole offers a complementary perspective on degrees of freedom that are already accounted for in the exterior description.  This means that there are nonlocal identifications between interior and exterior degrees of freedom.  Interior and exterior operators do not commute, violating locality across the horizon.

For an outside observer, black hole complementarity makes the interior of the black hole redundant, so that the space-time effectively ends at the horizon.  He sees anything thrown into the black hole approach the horizon, but not cross it.  Instead, infalling material is incinerated in a thermal layer near the horizon and gradually reemitted as nearly thermal Hawking radiation.
Since the black hole reaches thermal equilibrium quickly in the outside view, 
the entropy of the black hole, $S\sub{BH}$, is maximal and reflects the number of accessible states consistent with the black hole's mass, angular momentum and charge, $S\sub{BH} = \ln \Omega\sub{BH}$. We do not know the nature of the microscopic degrees of freedom that give rise to these states, but $\Omega\sub{BH}$ represents \emph{all} of the states of the black hole.  There are no additional states reflecting the internal state of the black hole, because there are no independent interior degrees of freedom.

\section{Excess interior entropy necessitates a stronger complementarity}\label{entropy}

The observer who falls into the black hole sees something quite different from the external observer.  While she is outside the horizon she also sees the thermal horizon layer.  As she falls toward the horizon, the thermal layer moves into the black hole well ahead of her.  If the black hole is very large she will notice nothing out of the ordinary when she crosses the horizon.  She can remember her past, perform local experiments, and admire the distant stars, even as the outside observer believes that she is being incinerated in the horizon boundary layer. 
In her view, she approaches the thermal layer only as she approaches the singularity.  

While the experiences of the inside and outside observers seem totally contradictory, black hole complementarity tells us that the observations are not only compatible, but actually represent complementary perspectives on the same degrees of freedom.  The identification between the interior and exterior descriptions is very complicated and not known, so an outside observer inspecting the horizon layer would not be able to determine what the inside observer is doing as she falls toward the singularity.

While the identifications between states is complicated, the fact that they are identified means that the interior description must have the same number of states as the exterior description, $\Omega\sub{BH}$.  In the interior description, objects are not cooked to equilibrium, so the entropy is not the maximum allowed by the number of states.  Therefore,
\begin{equation}
	S\sub{int} \leq S\sub{BH}\,.
	\label{eq:Bound}
\end{equation}
This inequality can be checked in any particular case by calculating the entropy seen in the interior description.  In the companion paper \cite{Wallace:2008zz}, we calculated the entropy created inside an accreting, electrically charged black hole.\footnote{Charged black holes are not thought to be realistic.  However, real black holes are expected to have significant angular momentum.  Since charged black holes have many of the same features as spinning black holes, while being easier to study, charge is often used as a surrogate for angular momentum.}  As the accreted, conducting matter falls toward the singularity, the black hole's electric fields create strong currents resulting in extravagant entropy creation.
Adding this entropy over space-like time slices inside the black hole violates bound (\ref{eq:Bound}) by an enormous factor.

Just as in section \ref{thermo}, the entropy has been over-counted, pointing to a break down of locality inside black holes. 
This time the break down is much more dramatic than the one proposed by black hole complementarity.  Since the entropy exceeds $S\sub{BH}$ by a huge factor, there must be a proliferation of complementary descriptions of the black hole's interior, each description having less than the Bekenstein-Hawking entropy.

\EPSFIGURE{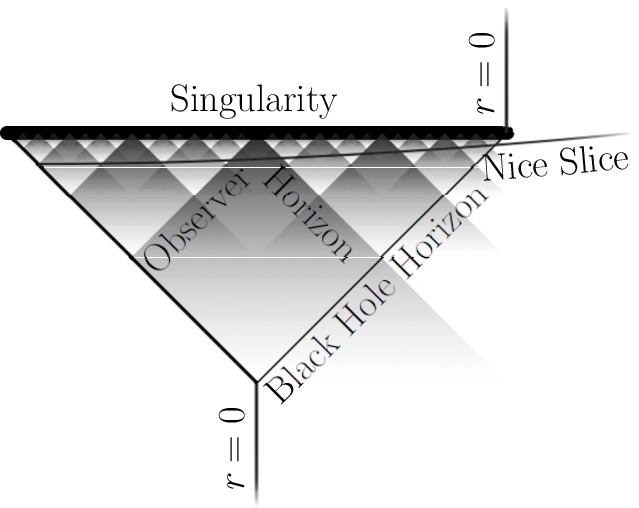}{The Penrose diagram of the black hole's interior.  Close to the singularity more observer horizons are required to cover the space-like slice.\label{Observers}}

Observer complementarity asserts that every observer sees the of the black hole's degrees of freedom \cite{Banks:2001yp}.  Since every observer in the black hole interior has her own observer horizon (figure \ref{Observers}), observer complementarity proposes an elaborate web of identifications in the black hole's interior, far more identifications than black hole complementarity.%
\footnote{Observer complementarity has been pursued primarily in deSitter spaces, where every observer is surrounded by a horizon that behaves much like a black hole horizon \cite{Banks:2001yp}.}
If observer complementarity is correct, then entropy should not be added across observer horizons, and only the the entropy seen by individual observers must be less than the Bekenstein-Hawking entropy.
Indeed, we will show in the next section that individual observers in the excess entropy example do not see an excess of entropy.  This offers strong support for observer complementarity.

\section{Observer complementarity solves the problem of excess interior entropy}\label{observer}

The model studied in the companion paper has matter falling into the black hole continuously.  Rather than finding the total entropy seen by the infalling observer, we calculate only the observed entropy, $\Delta S\sub{obs}$, of matter falling into the black hole in the vicinity of the infalling observer, shown in figure \ref{LightCone}.  This will be compared to the increase in the entropy of the black hole, $\Delta S\sub{BH}$, caused by the same matter as it falls through the horizon.
$\Delta S\sub{BH}$ is found from the increase in the horizon area seen by the outside observer, and is much larger than the entropy of the infalling matter as it falls through the horizon (by a factor of $10^{19}$ in the excess entropy example).

\DOUBLEFIGURE{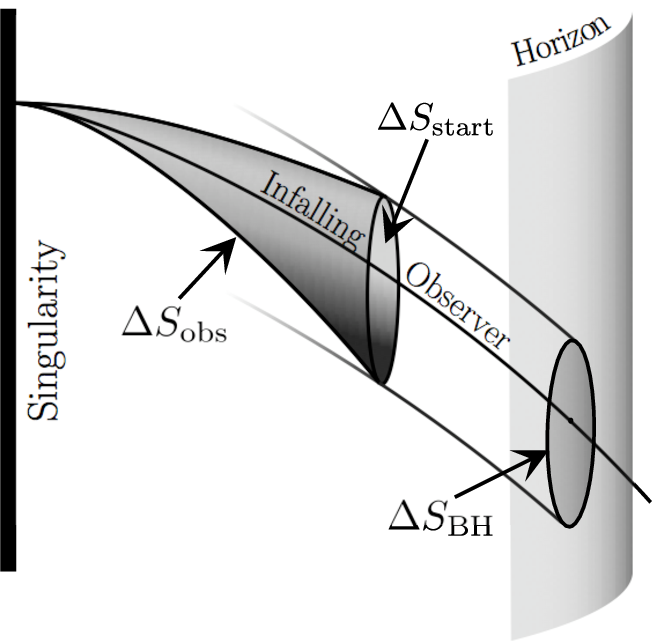}{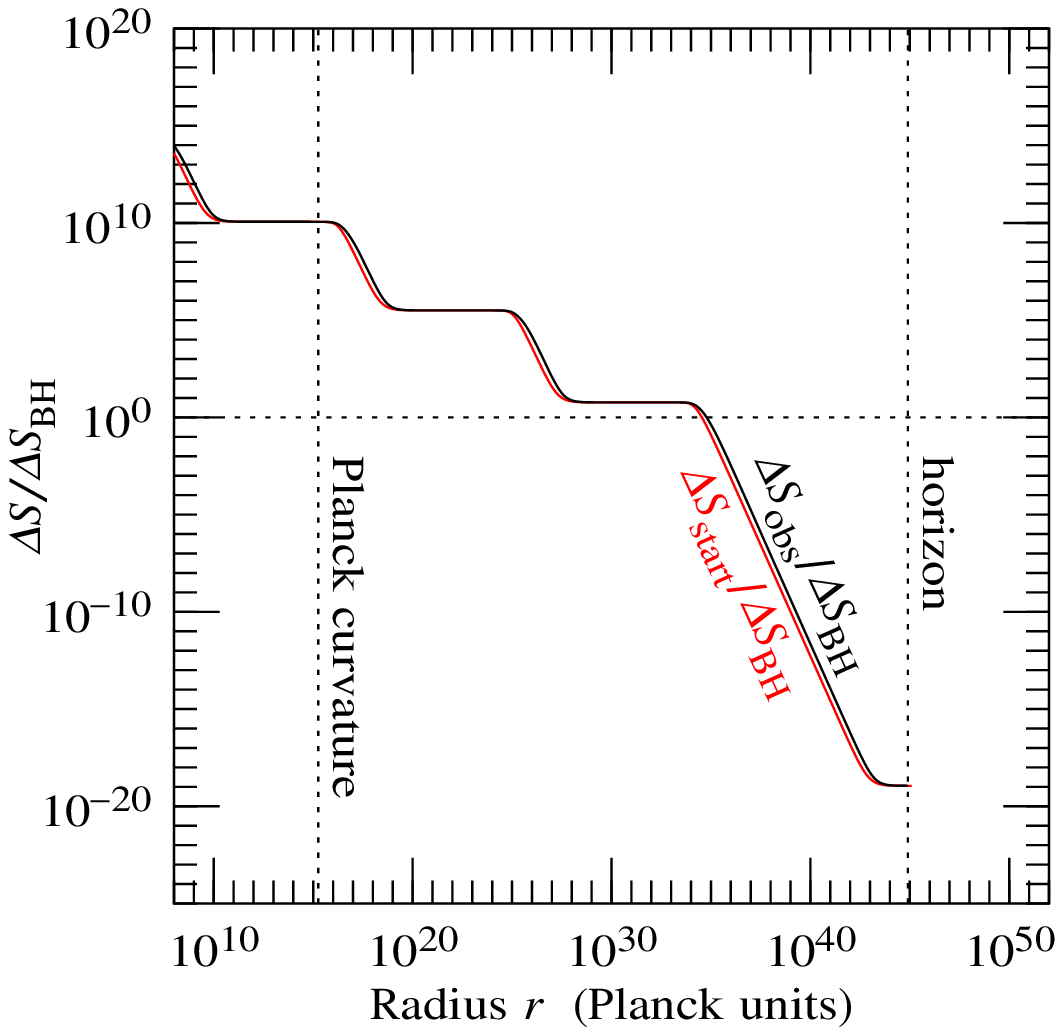, width=3in}{The observer horizon is the converging light cone that reaches the infalling observer just as she reaches the singularity.
\label{LightCone}}
{At any radius, the entropy that will pass through the observer horizon, $\Delta S\sub{obs}$, is only slightly greater than the entropy within the observer horizon, $\Delta S\sub{start}$.\label{EntropyGraph}}

The matter falls along with the infalling observer, heating and increasing in entropy.  Eventually this matter begins to leave the observer horizon.
The entropy per baryon diverges, but the volume inside the observer horizon shrinks sufficiently quickly for the total entropy seen by the infalling observer to be finite.
The calculation in appendix \ref{calculation} shows that $\Delta S\sub{obs}$ is less than $\Delta S\sub{BH}$.

The exact amount of entropy seen will depend on the size of the vicinity used in the calculation. 
If the vicinity is chosen so that matter starts leaving the observer horizon when it has an entropy $\Delta S\sub{start}$, then the total entropy seen by the infalling observer before she hits the singularity, $\Delta S\sub{obs}$, is of the same order of magnitude as  $\Delta S\sub{start}$.
Figure \ref{EntropyGraph} shows both $\Delta S\sub{start}/\Delta S\sub{BH}$ and $\Delta S\sub{obs}/\Delta S\sub{BH}$ as functions of the radius at which the matter starts to leave the horizon. 
If the vicinity is large enough that the matter starts leaving the observer horizon as soon as it goes through the black hole horizon, a reasonable starting point, then $\Delta S\sub{obs} \approx 10^{-19}\Delta S\sub{BH}$. 

Picking too small of a vicinity can give a value of $\Delta S\sub{obs}$ that is greater than $\Delta S\sub{BH}$, but this is of no physical significance.  One might hope to exceed the bound (\ref{eq:Bound}) by only dropping matter into the black hole in a tiny cloud near the infalling observer, so that none of it leaves until the entropy of the cloud has exceeded $\Delta S\sub{BH}$.  However, this situation is totally different from the continuous feeding of the black hole in the excess entropy example.  There is no reason to believe that the values of $\Delta S\sub{obs}/\Delta S\sub{BH}$ in figure \ref{EntropyGraph} would hold for the small cloud.  There are huge pressures compressing the matter in the excess entropy example.  Without those pressures the cloud would leave the observer horizon much earlier, greatly reducing $\Delta S\sub{obs}$.

The case considered in the companion paper is a bit unusual in that it has periodic self similarity, which results in the pulses of entropy production seen in figure \ref{EntropyGraph}.  Generic choices of conductivity do not produce pulses.  These generic cases also have $\Delta S\sub{obs} \ll \Delta S\sub{BH}$, as discussed in appendix \ref{calculation}.

\section{Conclusion}

Infalling observers see less than the Bekenstein-Hawking entropy, which is consistent with the number of degrees of freedom of the black hole.
Observer complementarity forbids adding entropy across horizons, so the problem of excess entropy inside black holes is avoided.

If an individual observer could see excess entropy then an even stronger set of identifications would be required.  No stronger complementarity has been proposed, and none is likely since the non-locality would be visible to the observer, a startling idea.

The excess entropy model is astrophysically realistic.  However, the limits on entropy should hold even for wildly unrealistic black holes.  It is possible the the Bousso bound guarantees that no observer can see more than the Bekenstein-Hawking entropy in any black hole, but we have not investigated this adequately to draw a conclusion on this point.

\acknowledgments
We would like to thank Don Page for his correspondence, which was both thought provoking and practical.  G. Polhemus would like to thank the Stanford Institute for Theoretical Physics for their helpful conversations and generous hospitality.  This work was supported in part by NSF award AST-0708607.

\appendix
\section{Calculation of visible entropy} \label{calculation}

We wish to know if the observed entropy, $\Delta S\sub{obs}$, is greater than the entropy increase of the black hole, $\Delta S\sub{BH}$.
In this section we calculate the ratio $\Delta S\sub{obs}/\Delta S\sub{BH}$ and find that it is much less than one in the excess entropy example \cite{Wallace:2008zz}, as well in examples with more generic conductivities.

The entropy seen by the infalling observer, $\Delta S\sub{obs}$, does not depend on her path, only on the time at which she hits the singularity, since this determines her observer horizon.  Therefore, we will assume she travels on the $z$-axis ($\theta = 0$), falling from the sonic point along with the ingoing baryons.  

\EPSFIGURE[t]{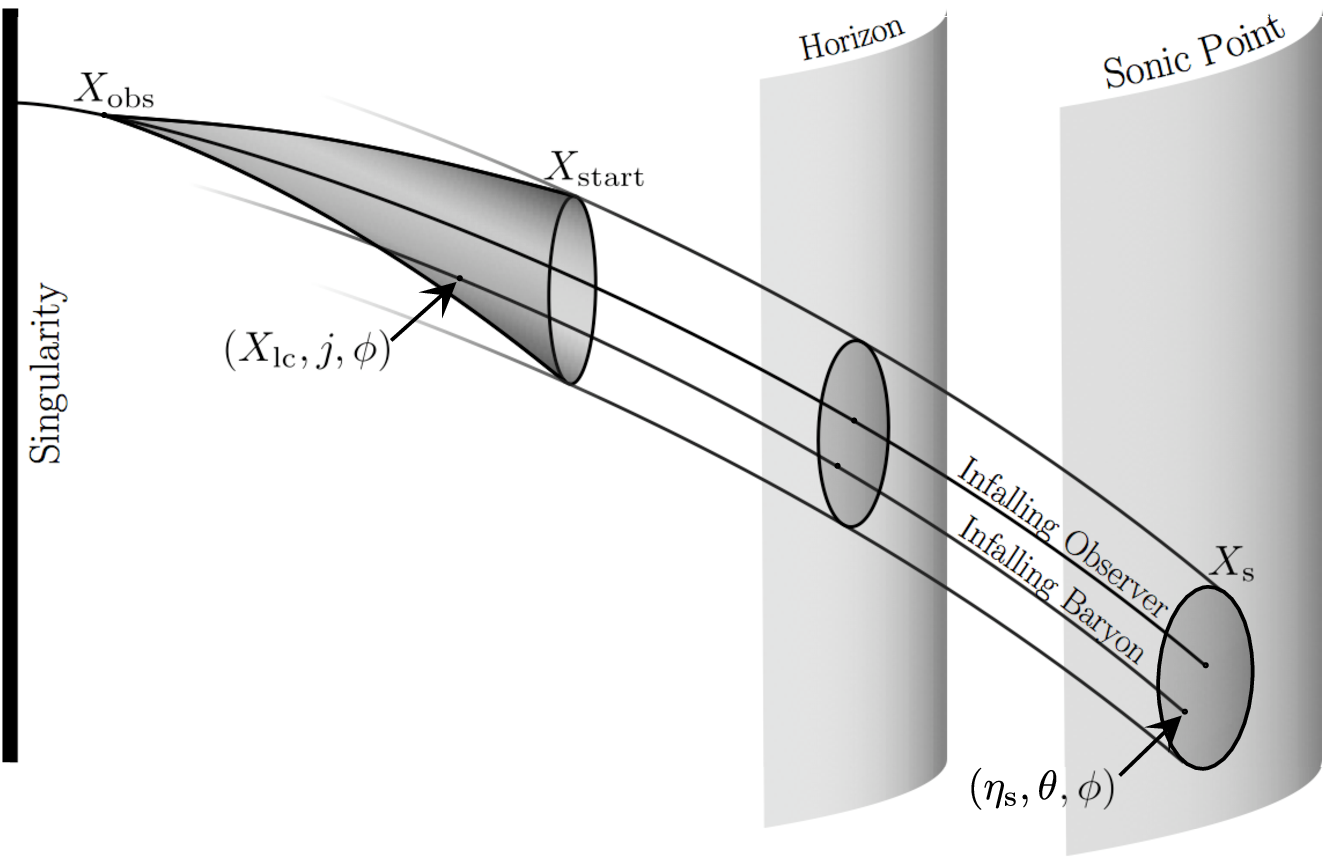}{An observer falls along the $\theta = 0$ axis.  A nearby baryon falls through the sonic point at conformal time $\eta\sub{s}$ and at angle $\theta$ from the axis.  Later, the baryon passes through the observer's past light cone at conformal radius $X\sub{lc}$ intersecting the path of a photon with angular momentum proportional to $j$.  The azimuthal angle $\phi$ of the baryon is the same as that of the photon.  The observer horizon is the observer's last past light cone, when $X\sub{obs}$ goes to the singularity.}

The baryons that fall into the black hole will also start from the sonic point.  The baryon trajectories are parameterized by the conformal time they leave the sonic point, $\eta\sub{s}$, and the spherical coordinates of the radial path that they follow into the black hole, $\theta$ and $\phi$.  The addition of these baryons will cause an increase in the horizon area of the black hole, thereby increasing the entropy seen by outside observers by $\Delta S\sub{BH}$ 
\begin{align}
	\Delta S\sub{BH}
	&= \int \frac{dS\sub{BH}}{d\eta\sub{s}}
		\frac{\sin\theta\, d\theta\, d\phi}{4\pi}\,d\eta\sub{s}\nonumber\\
	&= \frac{1}{2} \frac{dS\sub{BH}}{d\eta\sub{s}}
		\int\sin\theta\,d\theta\,d\eta\sub{s}	\,,
	\label{Sbh}
\end{align}
where $dS\sub{BH}/d\eta\sub{s}$ is the horizon entropy increase per conformal time as seen by an outside observer, which is constant over the times that we are considering (much less than the accretion time).  We divide this by $4\pi$ and integrate over the spherical angle of the region around the infalling observer to get the contribution of only those baryons in her vicinity.  Nothing depends on $\phi$, so the $\phi$ integral is done immediately.  The $\theta$ integral is left for later in order to accommodate a change of integration variables.

These baryons will fall along with the observer, heating all the way, until they reach her observer horizon.  To calculate the entropy that she will see, the entropy falling through the sonic point per conformal time, $dS\sub{BH}/d\eta\sub{s}$, is multiplied by the factor $dS\sub{lc}/dS\sub{s}$, representing the increase in entropy during the fall (the subscript ``lc'' signifies the entropy on the light cone that is her observer horizon):
\begin{align}
	\Delta S\sub{obs}
	&= \int \frac{dS\sub{lc}}{dS\sub{s}}\frac{dS\sub{s}}{d\eta\sub{s}}
		\frac{\sin\theta\, d\theta\, d\phi}{4\pi}\,d\eta\sub{s}\,.
\end{align}
The rate at which entropy falls through the horizon is constant over times much shorter than than the accretion time, so $dS\sub{s}/d\eta\sub{s}$ is constant and can be taken outside the integral:
\begin{align}
	\Delta S\sub{obs}
	&= \frac{1}{2}\frac{dS\sub{s}}{d\eta\sub{s}}\int \frac{dS\sub{lc}}{dS\sub{s}}
		\sin\theta\,d\theta\,d\eta\sub{s}\,.
	\label{Sobs}
\end{align}
The factor $dS\sub{lc}/dS\sub{s}$ depends on distance fallen by the baryons before they leave the horizon.  Only the baryons which fall into the black hole very close to the observer will have a huge increase in entropy before leaving the horizon.

Dividing equation (\ref{Sobs}) by equation (\ref{Sbh}) provides the desired ratio:
\begin{align}
	\frac{\Delta S\sub{obs}}{\Delta S\sub{BH}}
	&= \frac{dS\sub{s}}{dS\sub{BH}}
		\frac{1}{A}\int \frac{dS\sub{lc}}{dS\sub{s}}\sin\theta\,d\theta\,d\eta\sub{lc} \nonumber\\
	&= \frac{1}{A}\int \frac{dS\sub{lc}}{dS\sub{BH}}
		\sin\theta\,d\theta\,d\eta\sub{s}\,,
	\label{ratio}
\end{align}
where 
\begin{equation}
	A =\int \sin\theta\,d\theta\,d\eta\sub{s}\,.
	\label{A1}
\end{equation}
The factor $dS\sub{lc}/dS\sub{BH}$ is computed numerically in the companion paper for the specific parameters of the excess entropy example.
The integrals in equations (\ref{ratio}) and (\ref{A1}) are most easily performed over the observer horizon rather than over baryon trajectories.  The observer's past lightcone has two halves, one consisting of ingoing photons and the other of outgoing photons.  Each half is composed of photon trajectories parameterized by $\phi$ and $j$ ($j$ is proportional to the angular momentum and will be defined below).  The parameter $j$ goes from zero, for photons falling along the $\theta=0$ axis, to infinity at the boundary between ingoing and outgoing.

The position along the photon trajectories is best parameterized by the dimensionless ``ray-tracing'' radial coordinate, $X$ \cite{Hamilton:2004av}.  Since  $X$ is dimensionless, it is constant at the sonic point and the horizon.  $X$ goes to $-\infinity$ at the singularity.  We wish to determine the amount of entropy seen by the infalling observer when she has reached $X\sub{obs}$, so we will perform the integral over the light cone whose vertex is at $X\sub{obs}$.  There is no need to integrate over the entire light cone, only over some part that catches a sufficient amount of the entropy near the infalling observer.  We will start counting the entropy at $X\sub{start}$ and find all of the entropy that passes through the light cone from there to $X\sub{obs}$.  If $X\sub{start}$ is inside the horizon, then we will be able to ignore turning points and other complications.  The resulting integral over the light cone is
\begin{align}
	\frac{\Delta S\sub{obs}}{\Delta S\sub{BH}}
	&= \frac{1}{A}\sum_{\substack{\text{ingoing}\\\text{outgoing}}}
		\int_{X\sub{start}}^{X\sub{obs}}
		\frac{dS\sub{lc}}{dS\sub{BH}}
		\left(\int_{0}^{\infinity}
		\sin\theta \abs{D} \,dj\right)dX\sub{lc}\,,
\end{align}
where $D$ is the Jacobian determinant for the change of integration variables and
\begin{align}
	A	&= \sum_{\substack{\text{ingoing}\\\text{outgoing}}}
		\int_{X\sub{start}}^{X\sub{obs}}
		\int_{0}^{\infinity}
		\sin\theta \abs{D} \,dj\,dX\sub{lc}\,.
\end{align}

The expressions for the baryon trajectory parameters, $\phi$, $\theta$ and $\eta\sub{s}$, must be found in terms of the light cone parameters, $\phi$, $j$ and $X\sub{lc}$.  The angle $\phi$ is the same for the infalling baryon and the light cone photon. 

The expression for $\theta$ requires a straightforward integration that depends on both $j$ and $X\sub{lc}$ \cite{Hamilton:2004av}
\begin{align}
	\theta &= \pm \int_{X\sub{obs}}^{X\sub{lc}}\frac{J\,dX}{\sqrt{\tau^{2}-HJ^{2}}} \nonumber\\
	&= j \int_{X\sub{obs}}^{X\sub{lc}}\frac{dX}{\sqrt{1-Hj^{2}}}\,,
\end{align}
where $H$ is the homothetic scalar and $\tau$ is the proper time, which is constant for a photon.  We therefore replace $J$ with $j=\pm J/\tau$.  The positive sign is for ingoing photons, and the negative for outgoing.

Notice that if $-H$ grows slower than $X^{2}$ for large $-X$, then the integral for $\theta$ would diverge, and photons would make an infinite number of orbits before hitting the singularity.  However, in our models $-H$ grows faster than $X^{2}$, and we will use this fact to justify several useful approximations in the calculation of the more general case.

Since $\eta\sub{s}$ does not appear in the integrand, we will need only its derivatives with respect to the light cone coordinates in order to find the Jacobian determinant.  It is found in a similar manner to $\theta$ \cite{Hamilton:2004av}
\begin{align}
	\eta\sub{s}
	 &= \eta\sub{obs}
	 	\pm \int_{X\sub{obs}}^{X\sub{lc}}\frac{dX}{H\sqrt{1-Hj^{2}}}
		- \int_{X\sub{lc}}^{X\sub{s}}\frac{dX}{\xi^{t}\xi^{r}}\,.
\end{align}
The last term is the change in $\eta$ along the world line of the infalling baryons.  It does not have a sign change because the baryons are always ingoing.

The derivatives required for the Jacobian are
\begin{align}
	\frac{\partial\theta}{\partial X\sub{lc}} &= 
		\frac{j}{\sqrt{1-H\sub{lc}j^{2}}}\,,\\
	\frac{\partial\theta}{\partial j}  &=
		\int_{X\sub{obs}}^{X\sub{lc}}\frac{dX}{(1-Hj^{2})^{\frac{3}{2}}}\,,\\
	\frac{\partial\eta\sub{s}}{\partial X\sub{lc}} &= 
	 	\frac{\pm 1}{H\sub{lc}\sqrt{1-H\sub{lc}j^{2}}}
		+ \frac{1}{\xi\sub{lc}^{t}\xi\sub{lc}^{r}}\,, \\
	\frac{\partial\eta\sub{s}}{\partial j} &=
		\pm j \int_{X\sub{obs}}^{X\sub{lc}}\frac{dX}{(1-Hj^{2})^{\frac{3}{2}}}\,.
\end{align}

The Jacobian determinant, $\abs{D}$, is
\begin{equation}
	\abs{D} = \left[\frac{\sqrt{1-H\sub{lc}j^{2}}}{-H\sub{lc}} \mp \frac{1}{\xi^{t}\xi^{r}} \right]
		\int_{X\sub{obs}}^{X\sub{lc}}\frac{dX}{(1-Hj^{2})^{\frac{3}{2}}}\, ,
\end{equation}
where we have used the fact that the first term in the brackets is much larger than the second in taking the absolute value.
The term proportional to $(\xi^{t}\xi^{r})^{-1}$ is positive for the outgoing half of the light cone and negative for the outgoing half, so it will cancel out in the sums.

The resulting equations are collected here:
\begin{align}
	\frac{\Delta S\sub{obs}}{\Delta S\sub{BH}}
		&= \frac{2}{A}
		\int_{X\sub{start}}^{X\sub{obs}}
		\frac{dS\sub{lc}}{dS\sub{BH}}
		\left(\int_{0}^{\infinity}
		\sin\theta \abs{D\sub{ave}} \,dj\right)dX\sub{lc}
		\label{Big} \,,\\
	A	&= 2
		\int_{X\sub{start}}^{X\sub{obs}}
		\int_{0}^{\infinity}
		\sin\theta \abs{D\sub{ave}} \,dj\,dX\sub{lc} 
		\label{A}\,,\\
	\theta &= \pm j \int_{X\sub{obs}}^{X\sub{lc}}\frac{dX_{1}}{\sqrt{1-H_{1}j^{2}}}\,, \\
	\abs{D\sub{ave}} &= \frac{\sqrt{1-H\sub{lc}j^{2}}}{-H\sub{lc}}
		\int_{X\sub{obs}}^{X\sub{lc}}\frac{dX_{2}}{(1-H_{2}j^{2})^{\frac{3}{2}}}\,.
\end{align}
These equations can all be integrated numerically using the techniques in the companion paper.  The result is shown in figure \ref{EntropyGraph}.  The numerical calculation can be trusted for $X\sub{start}$ inside the horizon.
Pushing $X\sub{start}$ outside the horizon introduces turning points and other challenges.  The location of $X\sub{obs}$ makes little difference as long as it is exceeds $X\sub{start}$ by a few orders of magnitude.

The conductivity in the excess entropy example was chosen to give the pulses of entropy production shown in figure \ref{EntropyGraph}.  More generic choices of the conductivity give a steady increase in entropy and the problem of finding the entropy seen by an individual observer can be addressed without resorting to numerical integration.  If the entropy production is excessive, then the integrals in equation (\ref{Big}) will diverge as $X\sub{obs}$ goes to $-\infinity$, approaching the singularity.

To see that they do not diverge, recall that $-H$ grows faster than $X^{2}$, so the integrals for $\theta$ and $D\sub{ave}$ are dominated by small $-X$.  The integrands are 1 until the integral is cut off at $-Hj^{2} = 1$.  Since $-X\sub{lc}$ is always smaller than $-X_{1}$ and $-X_{2}$, the cutoff in the $X_{1}$ and $X_{2}$ integrals also cuts off the $j$ integral at $H\sub{lc}j^{2} = 1$. 
Let $X_{j}$ be the value of $X$ where $-Hj^{2} = 1$.  Then
\begin{align}
	\theta &\sim j \int_{X_{j}}^{X\sub{lc}}dX_{1} \sim - j X_{j}\,,\\
	\abs{D\sub{ave}} &\sim \frac{X_{j}}{H\sub{lc}}\,.
\end{align}
Since only the matter near the infalling observer poses an excessive entropy threat, $\theta$ is small.
The $j$ integral in equations (\ref{Big})
becomes
\begin{align}
	\int_{0}^{\infinity} \sin(\theta) \abs{D\sub{ave}}\,dj
	&\sim -\int_{0}^{(-H\sub{lc})^{-\frac{1}{2}}} \hspace{-1em} 
		\sin( j X_{j})\frac{X_{j}}{H\sub{lc}}\,dj \nonumber\\
	&\sim -H\sub{lc}^{-1} \int_{0}^{(-H\sub{lc})^{-\frac{1}{2}}} \hspace{-1em} j X_{j}^{2}\,dj
	\label{generic}\,.
\end{align}
For generic values of the conductivity, $-H$ grows like $-X^{3}$, allowing us to find $X_{j} = j^{-2/3}$.  Putting this into (\ref{generic}) gives
\begin{align}
	\int_{0}^{\infinity} \sin(\theta) \abs{D\sub{ave}}\,dj
	&\sim -X\sub{lc}^{-3} \int_{0}^{(-X\sub{lc})^{-\frac{3}{2}}} \hspace{-1em} j^{-1/3}\,dj \nonumber\\
	&\sim X\sub{lc}^{-4}\,.
\end{align}
In order for the final integral over $X\sub{lc}$ to diverge, $\Delta S\sub{lc}/\Delta S\sub{BH}$ would have to grow at least as fast as $-X^{3}$.  However, the entropy grows only like $(-X)^{3/4}$, too slowly to cause a divergence.  The entropy that will be seen by an infalling observer is of the same order as the entropy inside his horizon at $X\sub{start}$, far less than $\Delta S\sub{BH}$.

%\nocite{*}

\bibliographystyle{JHEP}
\bibliography{BlackEigen}

\end{document}